\documentclass[letter]{aa}  

\newcommand{\sdss}{SDSS\,1212\,}
\usepackage{graphicx}
\usepackage{txfonts}
\usepackage{amstext}

\begin{document}

   \title{X-ray orbital modulation of a white dwarf accreting from an L dwarf}

  \subtitle{The system SDSS\,J121209.31+013627.7}

   \author{B. Stelzer
          \inst{1}
          \and D. de Martino \inst{2} 
          \and S. L. Casewell \inst{3} 
          \and G. A. Wynn \inst{3} \and M. Roy \inst{3}
          }

   \institute{INAF - Osservatorio Astronomico di Palermo,  Piazza del Parlamento 1,
  I-90134 Palermo, Italy \\
              \email{stelzer@astropa.inaf.it}
         \and
         INAF - Osservatorio Astronomico di Capodimonte, Salita Moiariello 16, I-80131 Napoli, Italy
         \and
         Department of Physics and Astronomy, University of Leicester, Leicester, LE1 7RH, UK
   }
         

   \date{Received XXX; accepted XXX}

 
  \abstract{In an {\em XMM-Newton} observation of the binary 
  SDSS\,J121209.31+013627.7, consisting of a white dwarf and an
L dwarf, we detect X-ray orbital modulation as proof of accretion
  from the substellar companion onto the magnetic white dwarf.  
  We constrain the system geometry (inclination as well as magnetic and pole-cap angle)    
  through modelling of the X-ray light curve, and we derive a mass accretion 
  rate of $3.2 \cdot 10^{-14}\,{\rm M_\odot/yr}$ from the 
  X-ray luminosity ($\sim 3 \cdot 10^{29}$\,erg/s). 
  From X-ray studies of L dwarfs, a possible wind driven from a 
  hypothesized corona on the substellar donor is orders of magnitude too weak to explain
  the observed accretion rate, while the radius of the L dwarf 
  is comparable to its Roche lobe ($0.1\,R_\odot$), 
making Roche-lobe overflow the likely accretion mechanism in this system. 
}


   \maketitle
%

\section{Introduction}\label{sect:intro}

Sensitive wide-field surveys such as SDSS and UKIDSS have boosted the number of known white dwarf (WD) main-sequence (MS) binaries. 
In the majority of these systems the non-degenerate component is an M dwarf. 

Over $2000$ systems consisting of a WD  and an M dwarf (henceforth WDMD binaries) 
are known \citep{Rebassa-Mansergas13.0}, but 
only about a dozen binaries that consist of a WD and an L dwarf (WDLD binaries) 
\citep[see Sect.~\ref{sect:discussion} for accreting systems and ][ for a summary of detached systems]{casewell14}. 
This might be due to the
difficulty of detecting such faint very late-type companions in the spectral
energy distribution (SED) of WDs. 

Similar to their higher-mass siblings, the WDMD binaries, the WDLD 
systems comprise both wide (separation tens to thousands of AU)
and close binaries (period $< 10$\,h). 
Only a handful of them are close systems 
(period $\sim $100\,min) in which the low-mass companion must have survived a 
common-envelope phase \citep{nordhaus13.0}. 
 The progenitors of wide systems composed of WD and brown dwarf (BD), the 
AU-scale BD-MS binaries, are also rarely discovered, and they define the
so-called `brown dwarf desert' \citep{Marcy00.0}, suggesting that systems composed
of WD and BD form rarely. However, the much more frequent  
WDMD binaries evolve into cataclysmic variables (CVs), and subsequently, both the binary separation
and the donor mass decrease 
over time, which converts the donor into a close-in substellar object. 

When the donor star leaves thermal equilibrium and starts to expand in response to its 
mass loss, the orbital evolution of a CV reverses. 
The period minimum that represents this change is observed and theoretically predicted
to be at $P_{\rm orb} \sim 82$\,min \citep{Gansicke09.0, Knigge11.0}, 
corresponding to a donor mass of $\sim 0.06\,M_\odot$. This means that the 
systems that have evolved beyond the period minimum,  
so-called period-bouncers, have a substellar donor. 
Binary population synthesis revealed that period-bouncing WDLD binaries are 
expected to be the greater portion ($\sim 70$\,\%) of the whole CV population
\citep{Howell01.1}. The predicted high abundance of such systems is 
at odds with the very low detection rate.  
Systems with a very low-mass companion are difficult to identify because the contrast 
between the WD and L dwarf at optical and infrared wavelengths
is low. 
Only a few binaries composed of a WD and a BD are confirmed 
spectroscopically. 

The type of binary interaction in close WD / MS systems is tied to the evolutionary state. 
During the CV phase, accretion occurs through 
Roche-lobe overflow, while in detached systems 
wind accretion may take place.
A handful of systems showing very weak mass accretion 
\citep[$<10^{-13}\,{\rm M_\odot/yr}$, e.g.][]{Schmidt07.0} were defined as 
low accretion rate polars \citep[LARPs;][]{Schwope02.0}. However,
in most of them the donor is underfilling its Roche lobe, meaning
that they are in a pre-CV phase, 
and they must be accreting from a wind \citep{Schwope09.0}.
Magnetic siphons that channel the entire wind of the donor onto the WD have been proposed
to explain this class of systems with a strongly magnetic WD but weak or absent X-ray
emission \citep{Webbink05.0}. 

Accretion rates of WD binaries can be inferred from the X-ray luminosity or from 
the broad-band UV to optical/near-IR spectral energy distribution (SED) arising from the
accretion disk. H$\alpha$ emission provides only an upper limit to the accretion rate,
since this line can also have a significant contribution from the donor star's 
chromospheric activity or its irradiation by the WD. For the same reason, detection of
X-rays at low levels ($L_{\rm x} \leq 10^{29}$\,erg/s) from WD binaries with M-dwarf  
donors is not sufficient to diagnose accretion.  However, magnetic activity drops sharply at
late-M spectral types \citep[e.g.][]{West04.1}, and the coronal X-ray emission 
levels of L dwarfs are generally below the current sensitivity limits 
\citep{Stelzer06.1}. Only one very nearby L dwarf has been detected
in X-rays so far, at a level of
$\log{L_{\rm x}}\,{\rm [erg/s]} = 25.4$ \citep{Audard07.2}. 
Therefore, an X-ray detection of a WDLD binary
clearly points at accretion, whether through wind or Roche-lobe overflow.
Only one short-period WDLD system has been detected as X-ray emitter so far,  
SDSS\,J121209.31+013627.7 (\sdss), 
which has been reported as a weak {\em Swift} source by \cite{Burleigh06.1}.

\sdss is a magnetic WDLD binary (spectral type DA + L5/L8) 
with an average field strength of $7$\,MG \citep{Schmidt05.0, Farihi08.0}. 
\cite{Schmidt05.0} found narrow H$\alpha$ emission with periodic radial velocity 
variation ($P \sim 90$\,min) and large amplitude 
consistent with an origin in the L dwarf's irradiated atmosphere. 
Photometric variability at the same period and roughly in anti-phase with the
H$\alpha$ emission also indicates that there is a hot spot on the WD surface
\citep{Burleigh06.1}.  The  $K_{\rm s}$ -band light curve was shown by 
\cite{Debes06.0} to be more consistent with cyclotron emission from a magnetic pole than
with irradiation of a companion star. The authors also noted a clear excess in the $H$ and 
$K_{\rm s}$ bands with respect to the expected SED of the WD, which
was spectroscopically confirmed by \cite{Farihi08.0} to be due to a  
late-L dwarf.  \citet{Burleigh06.1} determined that in an $11$\,ks
long {\em Swift} observation \sdss had a count rate of 
$2.6 \cdot 10^{-3}$\,cts/s. They fitted the {\em Swift} X-ray spectrum with various 
one-temperature models,   
but the accretion mechanism (wind vs Roche lobe) could not
be established. The photon statistics were also insufficient for studying the X-ray light curve.

Here we report on an {\em XMM-Newton} observation of \sdss from which we 
unambiguously confirm that it is an accreting WDLD system. We derive an improved
estimate for its mass accretion rate and discuss the result in the framework of 
evolutionary scenarios for this system.

\section{XMM-Newton observations and analysis}\label{sect:xmm}

{\em XMM-Newton} observed \sdss on 6 June 2015 for $22$\,ksec (Obs-ID 0760440101) 
with all EPIC instruments using the {\sc thin} filter and with the Optical Monitor 
(OM) in {\sc fast mode} using the $B$ band filter.

We restrict the analysis to EPIC/pn, which provides the highest sensitivity of the EPIC
detectors. The data analysis was carried out with {\em XMM-Newton}'s 
Standard Science Analysis System (SAS) version 15.0.0. The observation is not affected
by flaring particle background, therefore we used the full exposure time of $22$\,ksec for the analysis. 
We filtered the data for pixel patterns ({\sc 0 $\leq$ pattern $\leq$ 12}), 
quality flag ({\sc flag = 0}), and events channels ({\sc 200 $\leq$ PI $\leq$ 15000}).   
Source detection was performed in three energy bands:
$0.2-1.0$\,keV (S), $1.0-2.0$\,keV (M), and $2.0-12.0$\,keV (H) 
using a customized procedure
based on the steps implemented in the SAS task {\sc edetect\_chain}. 
The net EPIC/pn source count rates are given in Table~\ref{tab:table_pn}. 
We note that \sdss is also detected in the MOS cameras at a net count rate of 
$0.0167 \pm 0.0010$\,cts/s
and $0.0178 \pm 0.0010$\,cts/s in the broad band for MOS\,1 and MOS\,2, respectively. 

\begin{table}
\begin{center}
\caption{X-ray count rate and pulsed fraction ($PF_{\rm sine}$) from a sine fit
for EPIC/pn data in different energy bands.}
\label{tab:table_pn}
\begin{tabular}{rclcrr}\\ \hline
\multicolumn{3}{c}{Energy} & Band & net source rate & \multicolumn{1}{c}{$PF_{\rm sine}$} \\ 
\multicolumn{3}{c}{{\rm [keV]}} & label & [cts/s]   & \\ \hline
$0.2$ & \hspace*{-0.4cm}$-$ & \hspace*{-0.4cm}$12.0$  & $B$ & $0.0615 \pm 0.0021$ &  $1.0$ \\
$0.2$ & \hspace*{-0.4cm}$-$ & \hspace*{-0.4cm}$1.0$   & $S$ & $0.0323 \pm 0.0015$ &  $0.94 \pm 0.03$ \\
$1.0$ & \hspace*{-0.4cm}$-$ & \hspace*{-0.4cm}$2.0$   & $M$ & $0.0188 \pm 0.0012$ &  $0.93 \pm 0.05$ \\
$2.0$ & \hspace*{-0.4cm}$-$ & \hspace*{-0.4cm}$12.0$  & $H$ & $0.0104 \pm 0.0009$ &  $1.0$ \\
\hline
\end{tabular}
\end{center}
\end{table}

For the spectral and temporal analysis we allowed only pixel patterns with {\sc flag $\leq$ 4}.
We defined a circular photon extraction region with radius of $30^{\prime\prime}$ 
centred on the EPIC/pn source position. 
The background was extracted from an adjacent circular region with radius of 
$45^{\prime\prime}$ on the same CCD chip. 
The background subtraction of the light curve was carried out with the SAS task 
{\sc epiclccorr,} which also corrects for instrumental
effects. We then barycentre-corrected the photon arrival times
 using the SAS tool {\sc barycen}. 

\begin{figure}
\begin{center}
\includegraphics[width=7.7cm]{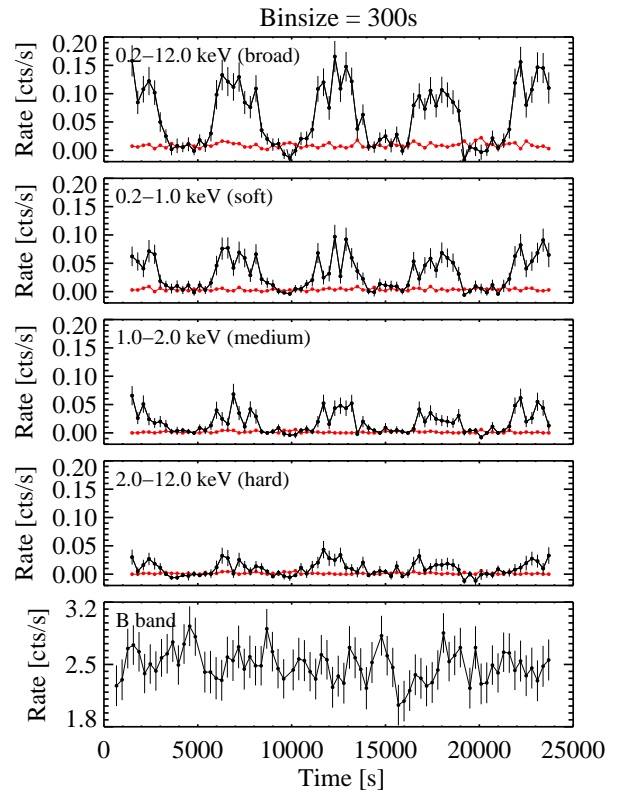}
\caption{OM $B$-band light curve and EPIC/pn X-ray light curve of \sdss 
with $1\,\sigma$ errors 
in four energy bands as labelled in the upper left of each panel. 
X-ray light curves represent the background-subtracted source signal (black) 
and for comparison the background signal (red). The bin size is $300$\,s.}
\label{fig:lcs_allenergies}
\end{center}
\end{figure}

The light curve shows a clear periodic modulation in all energy bands with larger
amplitude for softer emission (Fig.~\ref{fig:lcs_allenergies}). 
The modulation does not appear to be sinusoidal, displaying on-off behaviour that is
typical of strongly magnetized accreting WDs or AM\,Her systems \citep{Cropper90.0}.  
In the minimum the counts drop to zero, suggesting that the area of accretion, the pole cap,  
is completely occulted by the WD. 
All energy bands show flickering typical of X-ray emission from CVs. 
Lomb-Scargle periodogram analysis of the broad band ($0.2-12$\,keV) light curve yields
a period of 
$P_{\rm orb} = 88.3 \pm 0.6$\,min. 
This value and its $1\,\sigma$ error were derived with a bootstrap approach
from $5000$ simulated broad band light curves drawn randomly from the count rates and 
errors. This period is 
in good agreement with published periods for the H$\alpha$ emission 
\citep[$93.6 \pm 14.4$\,min;][]{Schmidt05.0}, 
the near-IR photometry \citep[$87.84 \pm 1.44$\,min;][]{Debes06.0},
and the optical photometry \citep[$88.428 \pm 0.001$\,min;][]{Burleigh06.1}.

To first approximation we fitted a sinusoid to the light curve from 
which we determined the pulsed fraction ($PF$) 
taking into account the uncertainties on y-offset and amplitude of the sine curve. 
The values obtained for the individual energy bands (see Table~\ref{tab:table_pn}) 
are consistent with the $100$\,\% PF observed in polars.

\begin{figure}[t]
\begin{center}
\includegraphics[width=6.5cm, angle=270]{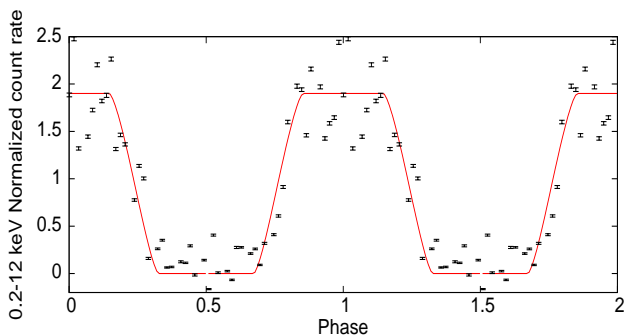}
\vspace*{-0.5cm}
\caption{
Barycentre-corrected X-ray (EPIC/pn $0.2-12$\,keV) light curve of
\sdss folded using the ephemeris of \cite{Burleigh06.1} and 
the period determined from the X-ray signal. 
The bin size is 450\,s. The best-fit model is shown in red.} 
\label{fig:lc_folded}
\end{center}
\end{figure}

The phase-folded X-ray light curve of \sdss is displayed in 
Fig.~\ref{fig:lc_folded} together with a geometric model based on 
the one presented by 
\citet{wynn1992} and \citet{brinkworth2004},
which describes direct accretion from a donor onto a WD magnetosphere 
with a post-shock region characterized through the angle between magnetic and 
rotation axis ($m$),  
the pole-cap opening half-angle ($b$), and the system inclination ($i$). 
We ran the model over a range of parameter space to determine the likely values for these 
angles, assuming a single pole to be responsible for the variation.
The modelling was performed on the X-ray light curve with a bin
size of $150$\,s. 
We found $80^\circ \gtrsim i \gtrsim 70^\circ$, 
$110^\circ \gtrsim m \gtrsim 100^\circ$, and $b \simeq 30^\circ$. 
These values meet 
the criteria for the self-occultation as in \citet{wynn1992} and references therein. 
We confirmed the size of the accretion region using different 
binnings with lower time-resolution. We also explored the effects of a vertical extent of the column up to $1\,R_{\rm WD}$ 
obtaining an upper limit of $0.2\,R_{\rm wd}$ and no major effect on the values 
of the other parameters. A small vertical extent of the emission region is also suggested
by the fact that the flux drops to zero and the ingress and the egress are steep. 

The EPIC/pn spectrum of \sdss was first fitted with single-component models: a power law,
a black body, or an optically thin model, none of which adequately describes the spectral
shape. 
When we add a simple absorber ({\sc tbabs}) and leave the abundance free to vary,
the thermal {\sc apec} model provides a reasonable fit. However, the $90$\,\% confidence
level of the abundance is compatible with zero, which is an unphysical result 
(see Table~\ref{tab:xspecfits}).
We also tested representations with two spectral components.  
In particular, the spectrum is compatible with 
an absorbed two-temperature ($2T$) thermal model
[{\sc tbabs $\cdot$ (apec + apec)}], which yields a 
better $\chi^2_{\rm red}$ than the one-temperature (1T) model.
However, similar to the case of the 1T-model, the abundance cannot be constrained 
by the fit.
If, in turn, the abundance is fixed to the value obtained from the 1T-model, 
the additional low-temperature component turns out to be completely unconstrained.
We conclude that the observed spectrum does not provide information on a possible
multi-temperature environment.  
There is therefore no evidence for an additional lower temperature blackbody component 
such as the one 
typically used to represent a soft excess in X-ray spectra of polars
and intermediate polars  
\citep[$<<0.1$\,keV, see e.g.][]{Beuermann12.0, Bernardini12.0}. Such a component would
also be 
physically unacceptable as it would be locally super-Eddington. 

%
\begin{table}
\begin{center}
\caption{Best-fit parameters for the EPIC/pn spectrum of \sdss 
and values corresponding to upper and lower $90$\,\% confidence ranges.}
\label{tab:xspecfits}
\begin{tabular}{lrrrr} \hline
\multicolumn{5}{c}{{\sc tbabs $\cdot$ apec}} \\  
 & $N_{\rm H}$ & $kT$  & & $Z$ \\
$\chi^2_{\rm red}$ (dof)   & [${\rm cm^{-2}}$] & [keV] & & [$Z_\odot$] \\ \hline
$1.13 (22)$            &  $2.3^{5.1}_{0.0} \cdot 10^{20}$ &  $2.62^{3.58}_{1.99}$ & &  $0.11^{0.44}_{0.0}$ \\ \hline
\end{tabular}
\end{center}
\end{table}

\begin{figure}
\begin{center}
\includegraphics[width=8.0cm]{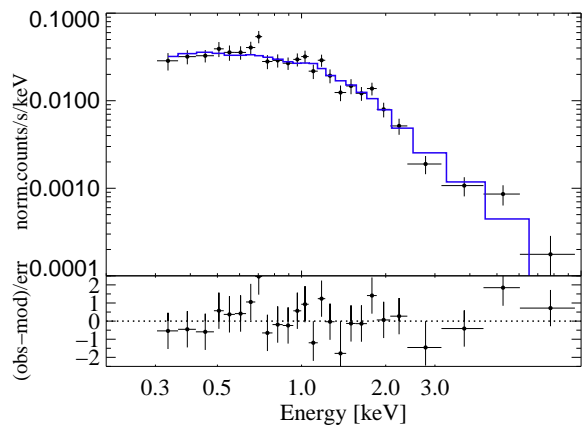}
\caption{Time-averaged EPIC/pn X-ray spectrum of \sdss with 
best-fitting one-temperature {\sc apec} model (solid blue line) and residuals.} 
\label{fig:xspec}
\end{center}
\end{figure}

We show in Fig.~\ref{fig:xspec} the observed EPIC/pn spectrum together with 
the 1T-model. 
The total galactic absorption in the direction of \sdss is 
$\sim 2 \cdot 10^{20}\,{\rm cm^{-2}}$ \citep{Dickey90.0}, 
consistent with the value determined for the EPIC/pn 
spectrum. Given the distance of $120$\,pc \citep{Burleigh06.1}, 
 the intrinsic absorption of the source must be very low. 

The X-ray flux is $1.5\,10^{-13}\,{\rm erg/cm^2/s}$, taking into account the 
bolometric correction to the $0.001 - 100$\,keV range, and $1.3\,10^{-13}\,{\rm erg/cm^2/s}$
for the $0.2-12$\,keV range. This latter value is 
similar to the value obtained by \cite{Burleigh06.1} from the {\em Swift} spectrum
($1.2\,10^{-13}\,{\rm erg/cm^2/s}$) for an absorbed $1T$ fit. Despite the low
statistics of the {\em Swift} data, these measurements, taken about one decade apart, 
indicate that the X-ray source is rather stable in time. 
With the distance ($d = 120$\,pc)
from \cite{Burleigh06.1}, we determine the bolometric X-ray luminosity to 
$\log{L_{\rm x}}\,{\rm [erg/s]} = 29.4$. 
From the X-ray luminosity we derive 
a mass accretion rate of $\dot{M}_{\rm acc} \sim 2.6\,10^{-14}\,{\rm M_\odot / yr}$; 
where we have used a slightly higher WD mass than 
\cite{Burleigh06.1} ($0.8\,M_\odot$; see Sect.~\ref{sect:discussion} for a justification)
and a correspondingly smaller radius ($R_{\rm WD} = 7 \cdot 10^8$\,cm).   
This value for 
$\dot{M}_{\rm acc}$ only includes the kinetic energy converted into X-ray luminosity. 
The value is higher when contributions from cyclotron emission and luminosity of the
hot spot are considered. 

The X-ray count rate in the minimum of the light curve ($0.35 \leq \phi \leq 0.65$)
is $0.0046$\,cts/s with a standard deviation of $0.0086$\,cts/s. This 
corresponds to an upper limit of the X-ray luminosity.  
With the count-to-flux conversion factor derived from the time-averaged 
X-ray spectrum, we obtain $\log{L_{\rm x,min}}\,{\rm [erg/s]} = 27$.
This value can be understood to represent the upper limit to any possible residual emission. 

$B$ -band photometry acquired with the OM in {\sc fast mode}, simultaneously with the X-ray observations, was extracted with the SAS
task {\sc omfchain,} and the time series was barycentre corrected. 
A Lomb-Scargle periodogram analysis did not yield a significant periodicity, and when 
folded on the X-ray period, no phase-related variability is seen (Fig.~\ref{fig:lcs_allenergies}, lowest panel).
At the time of the OM observation, \sdss was at $B_{\rm OM}$=18.28$\pm$0.08\,mag, which is 
consistent with the SED provided by \cite{Debes06.0} and with the 
$u^\prime$ and $g^\prime$ photometry shown by \cite{Burleigh06.1}. 
The optical light curve presented by \cite{Burleigh06.1} showed only a weak
modulation ($\lesssim 10$\,\% in $u^\prime$ and $\sim 4$\,\% in $g^\prime$). 
Given the low statistics of the OM data, it is therefore not surprising 
that no significant variability is detected in our $B$-band light curve.

\section{Discussion}\label{sect:discussion}

To date, only four period-bounce candidates have infrared detections:
EF\,Eri \citep{Schwope07.0}, SDSS\,J 143317.78+101123.3 \citep{Littlefair13.0, hernandez16}, 
WZ\,Sge \citep{Harrison16.0}, and \sdss \citep{Farihi08.0}. 
SDSS\,J 1433+1011 and WZ\,Sge are non-magnetic CVs, while EF\,Eri has been the textbook example for a LARP, but is also considered a candidate period
bouncer \citep{Schwope07.0, Schwope10.0}. 
The X-ray luminosity of EF\,Eri \citep[$2 \cdot 10^{29}$\,erg/s; see][]{Schwope07.0}
is remarkably similar to our measurement for \sdss. However, no X-ray variability has
be detected for EF\,Eri during its extended low state, implying that accretion (almost)  
stopped. Our detection of X-ray orbital modulation in \sdss, consistent with
its binary period found with other methods, provides unambiguous 
proof for accretion from 
the L dwarf onto the WD and establishes this system as a new benchmark for 
interacting binaries consisting of a WD and a substellar companion 
with low accretion rates. 

The X-ray luminosity of \sdss, both during our {\em XMM-Newton} observation in 
2015 and during the$\text{}$\,  {\em Swift} observation nine
years before, 
is several orders of magnitude higher than that of
any single L dwarf \citep[$< 10^{26}\,{\rm erg/s}$; 
see Audard et al. 2007 and the compilation by][]{Cook14.0}, 
clearly ruling out coronal emission from the ultracool companion.
Together with its orbital modulation, the high X-ray luminosity of \sdss therefore is 
another clear piece of evidence for accretion onto the magnetized WD. 

Following the lines of argument in \cite{Webbink05.0}, the mass flux driven by coronal X-ray
emission from an L dwarf with an assumed $\log{L_{\rm x}}\,{\rm [erg/s]} \sim 25$ would
be $10^{-16}\,{\rm M_\odot/yr}$ at most. 
The detection of winds from late-type stars requires high-resolution
UV spectroscopy and is at present not possible for L dwarfs. A very small number
of late-type stars have measured wind accretion rates, including two M dwarfs that show
much smaller $\dot{M}_{\rm w}$ than expected from the empirical $\dot{M}_{\rm w}$ vs $F_{\rm x}$
relation for solar-type stars 
\citep{Wood15.0}. Extrapolating from 
that relation to low $F_{\rm x}$, for an adopted $\log{L_{\rm x}}\,{\rm [erg/s]} \sim 25$ 
for the L dwarf, 
its wind would be $\approx 10^{-17}\,{\rm M_\odot/yr}$, making it difficult 
to provide the observed accretion rate of \sdss (considering 
that we measured a lower limit of $3 \cdot 10^{-14}\,{\rm M_\odot/yr}$).  
This leaves Roche-lobe overflow as the most likely
origin for the mass transfer. 

\citet{Farihi08.0} estimated the L-dwarf radius as
$0.09\,R_\odot$ and calculated the Roche lobe to be $R_{\rm L} = 0.11\,R_{\odot}$. 
However, this is based on the assumption that the WD mass is 
$0.6\,M_\odot$, a value typical of a non-magnetic, isolated WD. 
There is evidence that WDs in CVs
have higher masses than single WDs and those in detached systems \citep{zorotovic11}. 
Moreover, single magnetic WDs have also been found to be more massive than non-magnetic WDs 
\citep{Ferrario15.0}.
Considering that the non-magnetic CV SDSS\,1433+1011 -- with an
L-dwarf companion -- has a mass of 
$0.868\,M_{\odot}$ (\citealt{littlefair08}, \citealt{hernandez16}), assuming a mass of 
$\sim 0.8\,M_\odot$ for \sdss is therefore plausible. 
For this value, using Eq.~5 from \citet{Breedt12.0}, we determine 
$R_{\rm L} = 0.10\,R_{\odot}$, which is closer to the L-dwarf radius. 
Our detection of accretion, and thus the presence of an X-ray emitting region 
on the WD, also provides an explanation for the H$\alpha$ emission previously observed 
from \sdss, as due to irradiation of the donor by the X-ray emission from the accreting WD. 
As a result of the irradiation the L dwarf may be inflated, making Roche-lobe filling
even more likely. 

To conclude, \sdss presents the characteristics of the long-sought class of
period-bounce CVs (cool WD, substellar donor mass, and weak accretion), and it is
located in the observationally still nearly unpopulated 
`boomerang' region of CV evolution models \citep[e.g.][]{Howell01.1}
that awaits observational confirmation. 
\sdss is the first WDLD binary found to exhibit clear 
accretion-induced X-ray variability at a very low accretion rate. 
The WDLD binary EF\,Eri has long been known to show X-ray orbital 
modulation during high states \citep[e.g.][]{Patterson81.0}, 
but in its current low state
the X-ray emission could not be found to be modulated at the WD spin/binary orbit 
\citep{Schwope07.0}. 

The X-ray orbital modulation of \sdss 
suggests that accreting WDLD systems may be 
easy to identify in the X-ray band through 
the magnetically confined accretion flow onto the polar regions of the WD.
Sensitive searches for X-rays from WDLD binaries can therefore be expected to
provide further examples of pulsed emission, ensuing determination of mass accretion
rates at very weak levels, and the characterization of the multi-wavelength properties of
the so far widely elusive class of period-bounce CVs.

\begin{acknowledgements}
DDM acknowledges financial support from ASI INAF I/037/12/0. SLC acknowledges support
from the University of Leicester College of Science and Engineering. 
\end{acknowledgements}

\bibliographystyle{aa} 
\bibliography{sdss1212}

\end{document}